\documentclass[a4paper,12pt]{article}
\pdfoutput=1
\usepackage{amssymb}
\usepackage{tipa}
\usepackage{mathrsfs}
\usepackage{bbm}
\usepackage{epsf,amsmath}
\usepackage[dvips,usenames]{color}
\usepackage{graphicx,subfigure}
\usepackage{color}
\usepackage{colortbl}
\usepackage{booktabs,multirow,makecell}
\usepackage{cite}

\newlength{\dinwidth}
\newlength{\dinmargin}
\allowdisplaybreaks
\def\be{\begin{equation}}
\def\ee{\end{equation}}
\def\ba{\begin{align}}
\def\ea{\begin{align}}

\begin{document}

\title{\bf $\boldsymbol{B_s^0-\bar{B}_s^0}$ mixing within minimal flavor-violating two-Higgs-doublet models}
\author{Qin Chang$^{a,c}$, Pei-Fu Li$^{a}$ and Xin-Qiang Li$^{b,c}$\footnote{Corresponding author: xqli@itp.ac.cn}\\
{ $^a$\small Institute of Particle and Nuclear Physics, Henan Normal University, Henan 453007, P.~R. China}\\
{ $^b$\small Institute of Particle Physics and Key Laboratory of Quark and Lepton Physics~(MOE),}\\[-0.2cm]
{     \small Central China Normal University, Wuhan, Hubei 430079, P~R. China}\\
{ $^c$\small State Key Laboratory of Theoretical Physics, Institute of Theoretical Physics,}\\[-0.2cm]
{     \small Chinese Academy of Sciences, P.~R. China}
}

\date{}
\maketitle

\begin{abstract}
{\noindent}In the ``Higgs basis" for a generic 2HDM, only one scalar doublet gets a nonzero vacuum expectation value and, under the criterion of minimal flavor violation, the other one is fixed to be either color-singlet or color-octet, which are named as the type-III and type-C models, respectively. In this paper, the charged-Higgs effects of these two models on $B_s^0-\bar{B}_s^0$ mixing are studied. Firstly, we perform a complete one-loop computation of the electro-weak corrections to the amplitudes of $B_s^0-\bar{B}_s^0$ mixing. Together with the up-to-date experimental measurements, a detailed phenomenological analysis is then performed in the cases of both real and complex Yukawa couplings of charged scalars to quarks. The spaces of model parameters allowed by the current experimental data on $B_s^0-\bar{B}_s^0$ mixing are obtained and the differences between type-III and type-C models are investigated, which is helpful to distinguish between these two models.
\end{abstract}


\newpage

\section{Introduction}
\label{sec:intro}

Thanks to the successful running of the Large Hadron Collider~(LHC), particle physics has entered a new era, which is featured by the discovery of a new boson with a mass close to $125$~GeV~\cite{ATLASh,CMSh}. Its measured properties are so far in good agreement with those of the Standard Model~(SM) Higgs~\cite{ATLASh2,CMSh2,CDFD0h2}, suggesting that the electro-weak symmetry breaking~(EWSB) is probably realized via the Higgs mechanism implemented through a single scalar doublet. It should be noted, however, that the EWSB is not necessarily induced by just a single scalar. Interestingly, many new physics~(NP) scenarios are equipped with an extended scalar sector. The search for additional scalars is one of the important programs of the LHC experiments.

One of the extensions of SM scalar sector is the so-called two-Higgs-doublet model~(2HDM)~\cite{Lee2HDM}, in which a second  scalar doublet is added to the SM field content. To avoid the experimental constraints on flavor-changing neutral-current~(FCNC) transitions, which are forbidden at tree level in the SM due to the GIM mechanism~\cite{GIM}, two different hypotheses, natural flavor conservation~(NFC)~\cite{NFC} and minimal flavor violation~(MFV)~\cite{MFV}, have been proposed~\footnote{The NFC and MFV hypotheses are not the only alternatives to avoid constraints from FCNCs; models with controlled FCNCs have also been addressed in the literature and shown to be compatible with the data~\cite{FCNC_BGL}.}. In the NFC hypothesis, depending on the $Z_2$ charge assignments on the scalar doublets and fermions, there exist four types of 2HDM~(type-I, II, X and Y)~\cite{Branco}. In the MFV hypothesis, to control the flavor-violating interactions, all the scalar Yukawa couplings are assumed to be composed of the SM ones $Y^U$ and $Y^D$. In the ``Higgs basis"~\cite{Davidson:2005cw}, in which only one doublet gets a nonzero vacuum expectation value~(VEV) and behaves as the SM one, the allowed $SU(3)_C\otimes SU(2)_L\otimes U(1)_Y$ representation of the second scalar doublet that couples to quarks via Yukawa interactions is fixed to be either $(1,2)_{1/2}$ or $(8,2)_{1/2}$~\cite{Wise}, which implies that the second scalar doublet can be either color-singlet or color-octet. For convenience, they are referred to as type-III and type-C models~\cite{Slavich}, respectively. Examples of the former include the aligned 2HDM~(A2HDM)~\cite{A2HDM} and the four types of 2HDM reviewed in Ref.~\cite{Branco}. The scalar spectrum of the latter contains, besides a CP-even and color-singlet Higgs boson~(the usual SM one), three color-octet particles, one CP-even, one CP-odd and one electrically charged~\cite{Wise}.

Although the scalar-mediated flavor-violating interactions are protected by the MFV hypothesis, the type-III and type-C models still present very interesting phenomena in some low-energy processes, especially due to the presence of a charged Higgs boson~\cite{Wise,Slavich,Yuan}. In this paper, we shall study the $B_s^0-\bar{B}_s^0$ mixing within these two models and pursue possible differences between their effects. Since the charged Higgs contributes to the process at the same order as does the $W$ boson in the SM, the NP effects might be significant.

It is known that the $B_s^0-\bar{B}_s^0$ mixing, which is governed by a Schr\"odinger equation and induced by the $b\to s$ transition, plays an important role in accurate tests of the SM and indirect searches for NP. In terms of the off-diagonal elements of the mass and decay matrices, $M_{12}^s$ and $\Gamma_{12}^s$, the mass and width differences between the two mass eigenstates $|B_{H}\rangle$ and $|B_{L}\rangle$ are defined, respectively, by
\begin{align}\label{Dm-Dga-def}
\Delta M_s \equiv M_{H}-M_{L}=2 |M_{12}^s|\,, \quad
\Delta \Gamma_s \equiv \Gamma_{L}-\Gamma_{H}=2 |\Gamma_{12}^s|\cos\phi_s\,,
\end{align}
where $\phi_s \equiv \arg(-M_{12}^s/\Gamma_{12}^s)$ is the CP-violating phase. Such two observables have been measured precisely, with the averaged values given, respectively, by~\cite{HFAG}
\begin{align}
 \Delta M_s^{exp.}=17.757\pm0.021~{\rm ps^{-1}}\,,\quad
 \Delta \Gamma_s^{exp.}=0.081\pm 0.006~{\rm ps^{-1}}\,,
\end{align}
which are in good agreement with the recent SM predictions, $\Delta M_s^{SM}=(17.3\pm2.6)\,{\rm ps^{-1}}$ and $\Delta \Gamma_s^{SM}=(0.087\pm0.021)\,{\rm ps^{-1}}$~\cite{Lenz3}.

There are another two interesting observables related to $B_s^0-\bar{B}_s^0$ mixing, the flavor-specific CP asymmetry $a_{sl}^s$ and the CP-violating phase $\phi_s^{c\bar{c}s}$~\footnote{The phase $\phi_s^{c\bar{c}s}$ appears in tree-dominated $b\to c\bar{c}s$ $B_s$ decays, such as $B_s\to J/\psi\phi$, and is generally different from $\phi_s$ unless the terms proportional to $V_{cb}V_{cs}^{\ast}V_{ub}V_{us}^{\ast}$ and $(V_{ub}V_{us}^{\ast})^2$ in $\Gamma^s_{12}$ are neglected~\cite{Lenz1}.}, which are defined, respectively, by
\begin{equation}
a_{sl}^{s}={\rm Im}\frac{\Gamma^s_{12}}{M_{12}^s}=\frac{\Delta M_s}{\Delta \Gamma_s}\tan\phi_s\,, \qquad
\phi_s^{c\bar{c}s}={\rm arg}(M_{12}^s)\,.
\end{equation}
For $\phi_s^{c\bar{c}s}$, the SM prediction $\sim -0.036$~\cite{Lenz3} agrees with the experimental data $-0.015\pm0.035$~\cite{HFAG} within $1\sigma$ error bar. For $a_{sl}^{s}$, on the other hand, the measurement $(-0.75\pm0.41)\%$~\cite{HFAG} is significantly different from the SM estimation $\sim {\cal O}(10^{-5})$~\cite{Lenz3}, even through they are in agreement with each other at $1.5\sigma$ level due to the large experimental error bars. Under the constraints of the above four observables, the allowed NP spaces could possibly be strictly reduced. So, in this paper, we shall evaluate the effects of charged Higgs in type-III and type-C 2HDMs on $B_s^0-\bar{B}_s^0$ mixing, and pursue possible differences between these two models.

Our paper is organized as follows. In Sec.~2, after a brief review of the 2HDMs under the MFV hypothesis, we perform a complete one-loop computation of the electro-weak corrections to the amplitudes of $B_s^0-\bar{B}_s^0$ mixing within the two models. In Sec.~3, the numerical results and discussions are presented in detail. Finally, our conclusions are made in Sec.~4. Explicit expressions for the loop functions appearing in $B_s^0-\bar{B}_s^0$ mixing are collected in the appendix.

\section{Theoretical Framework}
\label{sec:theor}

\subsection{Brief review of the 2HDMs under the MFV hypothesis}
\label{subsec:2HDM}

Firstly, for convenience and consistence, we shall give a brief review of the 2HDMs under the MFV hypothesis. In the ``Higgs basis'', the Yukawa interactions of the two Higgs fields $\Phi_1$ and $\Phi_2$ with quarks are given by~\cite{Wise,Slavich}
\begin{align}\label{L:Yukawa}
-\mathcal{L}_Y=\bar q_L^0 \tilde \Phi_1 Y^U u_R^0
    +\bar q_L^0 \Phi_1 Y^D d_R^0
    +\bar q_L^0 \tilde\Phi_2^{(a)}T_R^{(a)}\bar Y^U u_R^0
    +\bar q_L^0       \Phi_2^{(a)}T_R^{(a)}\bar Y^D d_R^0
    +\rm{h.c.}\,,
\end{align}
where $q_L^0$, $u_R^0$ and $d_R^0$ are the quark fields in the interaction basis; $T_R^{(a)}$ is the $SU(3)_C$ generator and determines the color nature~(color-singlet or color-octet) of the second scalar doublet; $Y^{U,D}$ and $\bar Y^{U,D}$ denote the Yukawa couplings and are generally complex $3 \times 3$ matrices in the quark flavor space. According to the MFV hypothesis, the transformation properties of the Yukawa coupling matrices $Y^{U,D}$ and $\bar Y^{U,D}$ under the quark flavor symmetry group $SU(3)_{Q_L}\otimes SU(3)_{U_R} \otimes SU(3)_{U_D}$ are required to be the same. This can be achieved by requiring $\bar Y^{U,D}$ to be composed of pairs of the matrices $Y^{U,D}$~\cite{Slavich}
\begin{align} \label{eq:Y}
  \bar Y^U&=A_u^\ast (1+\epsilon_u^\ast Y^UY^{U\dagger}+\dotsc)Y^U\,, \nonumber\\[0.2cm]
  \bar Y^D&=A_d(1+\epsilon_d Y^UY^{U\dagger}+\dotsc)Y^D\,,
\end{align}
where the ellipses denote trivial terms involving higher power of $Y^U Y^{U\dagger}$ and powers of $Y^D Y^{D\dagger}$.

After applying the SM unitary transformations to rotate the fermionic fields from the interaction to the mass-eigenstate basis, one can finally obtain the Yukawa interactions of charged Higgs bosons with quarks in the mass-eigenstate basis~\cite{Slavich,Yuan}
\begin{align}\label{eq:LCH}
  \mathcal L_{H^+}=\frac{g}{\sqrt{2}m_W}\,\sum_{i,j=1}^3\,\bar u_iT_R^{(a)}(A_u^i m_{u_i}P_L-A_d^im_{d_j}P_R)V_{ij}d_jH_{(a)}^+ +\rm{h.c.}\,,
\end{align}
where $g$ is the $SU(2)_L$ coupling constant, $i\,,j$ the fermionic generation indices, and $m_{u,d}$ the quark masses; $V$ denotes the Cabibbo--Kobayashi--Maskawa~(CKM) matrix~\cite{Cabibbo:1963yz,Kobayashi:1973fv}, and $P_{R,L}=\frac{1\pm \gamma_5}{2}$ are the right- and left-handed chirality projectors. The couplings $A_{u,d}^i$ are generally family-dependent and read
\begin{align}
  A_{u,d}^i=A_{u,d}\left(1+\epsilon_{u,d}\frac{m_t^2} {v^2}\delta_{i3}\right),
\end{align}
where $v=\langle \Phi_1^0\rangle=174~{\rm GeV}$. Since only the couplings of charged Higgs bosons to the top quark are involved for $B_s^0-\bar{B}_s^0$ mixing, we shall drop the family index in $A_{u,d}^i$ from now on.

Following the notation used in Ref.~\cite{Slavich}, we shall denote the model with the second scalar doublet being color-singlet and the one with the second scalar doublet color-octet as the type-III and the type-C model, respectively, both of which satisfy the principle of MFV. Their explicit contributions to the $B_s^0-\bar{B}_s^0$ mixing
will be presented in the next subsection.

\subsection{$\boldsymbol{B_s^0-\bar{B}_s^0}$ mixing within the SM and the 2HDMs with MFV}
\label{subsec:wcs}

%%%%%%%%%%%%%%%%%%%%%%%%%%%%%%%%%%%%%%%%
\begin{figure}[t]
\begin{center}
\subfigure[]{\includegraphics[width=3.6cm]{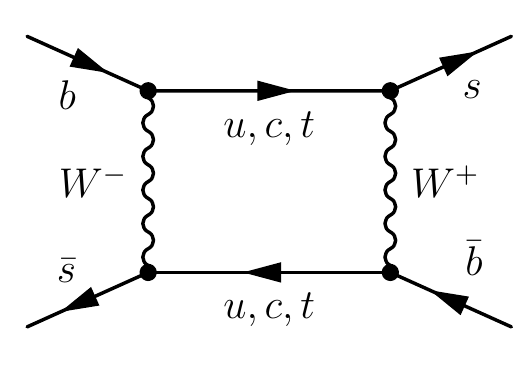}} \quad
\subfigure[]{\includegraphics[width=3.6cm]{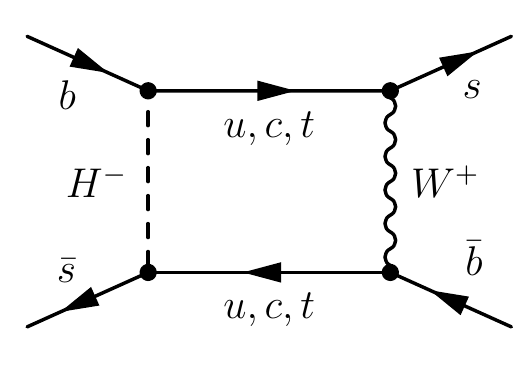}}
\quad
\subfigure[]{\includegraphics[width=3.6cm]{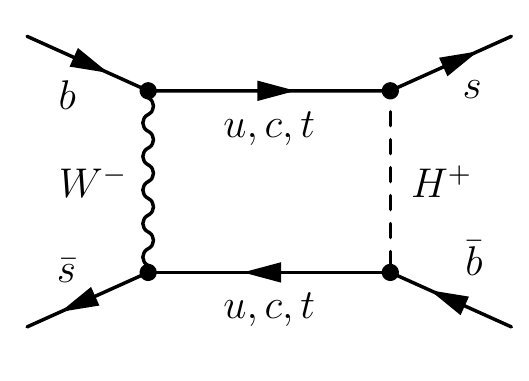}} \quad
\subfigure[]{\includegraphics[width=3.6cm]{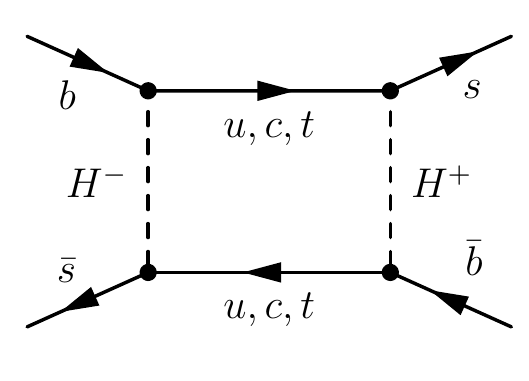}}
\caption{\label{topology} \small Box diagrams relevant to the $B_s^0-\bar{B}_s^0$ mixing in the unitary gauge, both within the SM~(the first one) and in the 2HDMs with MFV~(the last three ones). We have also taken into account the crossed diagrams, which are related to the original ones by interchanging the external lines.}
\end{center}
\end{figure}
%%%%%%%%%%%%%%%%%%%%%%%%%%%%%%%%%%%%%%%%

Within the 2HDMs with MFV, the $B_s^0-\bar{B}_s^0$ mixing occurs through the box diagrams shown in Figs.~\ref{topology}(b)-(d), which are obtained from the SM one~(Fig.~\ref{topology}(a)) with the $W^{\pm}$ propagator(s) replaced by the charged-Higgs $H^{\pm}$ one(s). After calculating these one-loop box diagrams and applying the standard procedure of matching~\cite{Buras:1990fn,Urban:1997gw}, one can obtain the 2HDM corrections to the $B_s^0-\bar{B}_s^0$ mixing. Together with the SM contribution, the resulting effective weak Hamiltonian responsible for $B_s^0-\bar{B}_s^0$ mixing can be written as
\begin{equation}\label{eq:heff}
  \mathcal{H}_{eff}^{full}=\frac{G_F^2}{16\pi^2}\,m_W^2\,\left(V_{tb}V_{ts}^*\right)^2\,\left[\,C^{VLL}(\mu) O^{VLL}+C^{SRR}(\mu)O^{SRR}+C^{TRR}(\mu)O^{TRR}\,\right]+\text{h.c.}\,,
\end{equation}
where $G_F$ is the Fermi coupling constant, and $C_{i}(\mu)$ the scale-dependent Wilson coefficients of the four-quark operators $O_{i}$, which are defined, respectively, as
\begin{align}\label{oper}
  &O^{VLL}=\bar{s}^\alpha\gamma_\mu(1-\gamma_5)b^\alpha\,\bar{s}^\beta\gamma^\mu(1-\gamma_5)b^\beta\,,\nonumber\\[0.2cm]
  &O^{SRR}=\bar{s}^\alpha(1+\gamma_5)b^\alpha\,\bar{s}^\beta(1+\gamma_5)b^\beta\,,\nonumber\\[0.2cm]
  &O^{TRR}=\bar{s}^\alpha\sigma_{\mu\nu}(1+\gamma_5)b^\alpha\,\bar{s}^\beta\sigma^{\mu\nu}(1+\gamma_5)b^\beta\,,
\end{align}
with $\alpha,\beta$ being the color indices and $\sigma_{\mu\nu}=\frac{1}{2}[\gamma_{\mu},\gamma_{\nu}]$. In addition, the hadronic matrix elements of the four-quark operators can be parameterized as~\cite{BurasDF2}
\begin{align}
  &\langle B_s^0 | O^{VLL} |\bar{B}_s^0 \rangle=\frac{8}{3}m_{B_s}^2f_{B_s}^2B_1(\mu_b)\,,\nonumber\\[0.2cm]
  &\langle B_s^0 |  O^{SRR} |\bar{B}_s^0 \rangle=-\frac{5}{3}\Big(\frac{m_{B_s}}{\bar{m}_b(\mu_b)+\bar{m}_s(\mu_b)}\Big)^2m_{B_s}^2f_{B_s}^2B_2(\mu_b)\,,\nonumber\\[0.2cm]
  &\langle B_s^0 |  O^{TRR} |\bar{B}_s^0 \rangle=\frac{4}{3}\Big(\frac{m_{B_s}}{\bar{m}_b(\mu_b)+\bar{m}_s(\mu_b)}\Big)^2m_{B_s}^2f_{B_s}^2\left[-5B_2(\mu_b)+2B_3(\mu_b)\right]\,,
\end{align}
where $f_{B_s}$ is the $B_s$-meson decay constant, and $B_i(\mu_b)$ are the non-perturbative bag parameters calculated on the lattice at a characteristic scale $\mu_b\sim\mathcal{O}(m_b)$~\cite{Carrasco:2013zta}.

The Wilson coefficients $C_i(\mu)$ in Eq.~(\ref{eq:heff}) consist of both the SM and 2HDM contributions and the results at the initial scale $\mu_W\sim\mathcal{O}(m_W,m_t,m_{H^{\pm}})$ can be written as
\begin{align}
  &C^{VLL}(\mu_W)=C^{VLL}_{SM}(\mu_W)+C^{VLL}_{2HDM}(\mu_W)\,,\nonumber\\[0.2cm]
  &C^{SRR}(\mu_W)=C^{SRR}_{SM}(\mu_W)+C^{SRR}_{2HDM}(\mu_W)\,,\nonumber\\[0.2cm]
  &C^{TRR}(\mu_W)=C^{TRR}_{SM}(\mu_W)+C^{TRR}_{2HDM}(\mu_W)\,.
\end{align}
Within the SM, the explicit expressions of the Wilson coefficients are computed from the box diagram shown in Fig.~\ref{topology}(a), accompanied by perturbative QCD corrections up to the desired order, details of which could be found, for example, in Refs.~\cite{Buras:1990fn,Urban:1997gw,Buchalla}. Including the next-to-leading order (NLO) QCD corrections, the SM contribution is given by~\cite{Buras:1990fn,Urban:1997gw,Buchalla}
\begin{equation}
  C^{VLL}_{SM}(\mu_W)=S_0(x_t)+\frac{\alpha_s(\mu_W)}{4\pi}\,\left[S_1(x_t)+F(\mu_W)S_0(x_t)+B_tS_0(x_t)\right]\,,
\end{equation}
where $x_t=\frac{\bar{m}_t^2(\mu_W)}{m^2_W}$, and the leading order (LO) coefficient $S_0(x_t)$ is the known Inami-Lim function~\cite{Inami:1980fz}. Generally, the SM also contributes to $C^{SRR}$ and $C^{TRR}$ and, in the absence of QCD corrections, we get
\begin{align}
  C^{SRR}_{SM}(\mu_W)\,=\,&\frac{x_b}{6}\bigg[\frac{x_t^2(5-22x_t+5x_t^2)}{3(1-x_t)^4}+\frac{x_t^2(1-3x_t-3x_t^2+x_t^3)\ln{x_t}}{(1-x_t)^5}\bigg]\,,\\[0.2cm]
  C^{TRR}_{SM}(\mu_W)\,=\,&\frac{x_b}{6}\bigg[-\frac{5-15x_t+8x_t^2-15x_t^3+5x_t^4}{3(1-x_t)^4}+\frac{(1-5x_t+9x_t^2-x_t^3)\ln{x_t}}{(1-x_t)^5}\bigg]\,,
\end{align}
with $x_b=\frac{\bar{m}_b^2(\mu_W)}{m^2_W}$. It is obvious that both $C^{SRR}_{SM}(\mu_W)$ and $C^{TRR}_{SM}(\mu_W)$ are suppressed by the factor $x_b$ and are, therefore, usually neglected in the literature~\cite{Buras:1990fn}.

The charged-Higgs contributions to the Wilson coefficients are computed from the last three box diagrams shown in Fig.~\ref{topology}, and depend on the two Yukawa coupling parameters $A_u$ and $A_d$, as well as the charged-Higgs mass $m_{H^{\pm}}$~\cite{Urban:1997gw,Buras:2001mb,Jung:2010ik,Manohar:2006ga,Cheng:2015lsa}. For the most general values of these parameters, especially when $A_d/A_u\simeq m_t/m_b$, each term in Eq.~(\ref{eq:LCH}) can give a comparable contribution and should be, therefore, taken into account simultaneously. Explicitly, for the color-singlet charged-Higgs contributions~(type-III model), we get
\begin{align} \label{eq:Ci_III}
  C^{VLL}_{III}(\mu_W)=&A_uA_u^*\,f_1(x_t,x_h)+A_u^2A_u^{*2}\,f_2(x_t,x_h)\,,\\[0.2cm]
  %%%%%
  C^{SRR}_{III}(\mu_W)=&-x_b\Big[A_uA_u^*\,f_3(x_t,x_h)+A_dA_u^*\,f_4(x_t,x_h)+A_d^2A_u^{*2}\,f_5(x_t,x_h)\,\nonumber\\[0.1cm]
                       &\hspace{1.1cm}+A_u^2A_u^{*2}\,f_6(x_t,x_h)+A_dA_uA_u^{*2}\,f_7(x_t,x_h)\Big]\,,\\[0.2cm]
  %%%%%
  C^{TRR}_{III}(\mu_W)=&0\,,
\end{align}
where $x_h=m^2_{H^{\pm}}/m^2_W$, and the explicit expressions for $f_i(x_t,x_h)$ are collected in the appendix. For the color-octet charged-Higgs contributions~(type-C model), on the other hand, we get
\begin{align} \label{eq:Ci_C}
  C^{VLL}_{C}(\mu_W)=&\frac{1}{3}\,A_uA_u^*\,f_1(x_t,x_h)+\frac{11}{18}\,A_u^2A_u^{*2}\,f_2(x_t,x_h)\,,\\[0.2cm]
  %%%%%
  C^{SRR}_{C}(\mu_W)=&-x_b\bigg[-\frac{5}{12}\,A_uA_u^*\,f_3(x_t,x_h)-\frac{5}{12}\,A_dA_u^*\,f_4(x_t,x_h)
                     -\frac{19}{72}\,A_d^2A_u^{*2}\,f_5(x_t,x_h)\,\nonumber\\[0.1cm]
                     &\hspace{1.2cm}-\frac{19}{72}\,A_u^2A_u^{*2}\,f_6(x_t,x_h) -\frac{19}{72}\,A_dA_uA_u^{*2}\,f_7(x_t,x_h)\bigg]\,,\\[0.2cm]
  %%%%%
  C^{TRR}_{C}(\mu_W)=&-x_b\bigg[\frac{1}{16}\,A_uA_u^*\,f_3(x_t,x_h)+\frac{1}{16}\,A_dA_u^*\,f_4(x_t,x_h)
                     +\frac{7}{96}\,A_d^2A_u^{*2}\,f_5(x_t,x_h)\,\nonumber\\[0.1cm]
                     &\hspace{1.2cm} +\frac{7}{96}\,A_u^2A_u^{*2}\,f_6(x_t,x_h) +\frac{7}{96}\,A_dA_uA_u^{*2}\,f_7(x_t,x_h)\bigg]\,.
\end{align}
It is noted that the Wilson coefficient $C^{TRR}_{C}(\mu_W)$ is now nonzero in the type-C model. To check the gauge independence of our results, we have performed the calculation both in the Feynman and in the unitary gauge. For $C^{VLL}$, our results agree with the ones presented in Refs.~\cite{Urban:1997gw,Buras:2001mb,Manohar:2006ga}~\footnote{There are two typos in Eq.~(26) of Ref.~\cite{Manohar:2006ga}: a global factor $2$ should be added to the term proportional to $|\eta_U|^2$ and $1/2$ to the term proportional to $|\eta_U|^4$.}. For $C^{SRR}$ and $C^{TRR}$, on the other hand, in order to get a gauge-independent result, the external momenta of the heavy quarks inside the mesons should be taken into account, and the heavy-quark masses should be kept up to the second orde; our results for these two coefficients differ from the ones presented in Refs.~\cite{Urban:1997gw,Buras:2001mb,Manohar:2006ga}.

The QCD renormalization group~(RG) evolution of these Wilson coefficients from the matching scale $\mu_W$ down to the lower scale $\mu_b$ has been calculated in Ref.~\cite{BurasDF2}. One can then obtain the corresponding Wilson coefficients at the scale $\mu_b$ through~\cite{BurasDF2}
\begin{align}
  C^{VLL}(\mu_b)&=\left[\eta(\mu_b)\right]_{VLL}C^{VLL}(\mu_W)\,, \\[0.2cm]
  \left(
      \begin{aligned}
         &C^{SRR}(\mu_b)\\[0.1cm]
         &C^{TRR}(\mu_b)
      \end{aligned}
   \right)&=
   \left(
      \begin{aligned}
         &\left[\eta_{11}(\mu_b)\right]_{SRR}\,\, &\left[\eta_{12}(\mu_b)\right]_{SRR}\\[0.1cm]
         &\left[\eta_{21}(\mu_b)\right]_{SRR}\,\,
         &\left[\eta_{22}(\mu_b)\right]_{SRR}
      \end{aligned}
   \right)
   \left(
     \begin{aligned}
         &C^{SRR}(\mu_W)\\[0.1cm]
         &C^{TRR}(\mu_W)
     \end{aligned}
   \right)\,,
\end{align}
where the explicit expressions of the evolution factors $\eta$ could be found in Ref.~\cite{BurasDF2}.

Equipped with the above information, the off-diagonal mass matrix element $M_{12}^s$ is given as
\begin{equation}
 M_{12}^s = \langle B_s^0|{\cal H}_{eff}^{full}|\bar{B}_s^0 \rangle
 ={\cal A}^{VLL}+{\cal A}^{SRR}+{\cal A}^{TRR}\,,
 \end{equation}
where ${\cal A}^{VLL}$, ${\cal A}^{SRR}$ and ${\cal A}^{TRR}$ denote the contributions induced by the three four-quark operators defined by Eq.~(\ref{oper}), respectively. Within the SM, the off-diagonal decay matrix element $\Gamma_{12}^s$ can be written as~\cite{Lenz,beneke}
\begin{eqnarray}
\Gamma_{12}^s(SM)=
    - \left[\, \lambda_t^2 \, \Gamma_{12}^{cc} \; + \;
        2 \, \lambda_t\,  \lambda_u \,
        \left( \Gamma_{12}^{cc} - \Gamma_{12}^{uc} \right) \; + \;
        \lambda_u^{2} \,
        \left( \Gamma_{12}^{cc} - 2 \Gamma_{12}^{uc} + \Gamma_{12}^{uu}\right)
       \right]\,, \label{ga12t}
\end{eqnarray}
with the CKM factors $\lambda_i=V_{ib}V_{is}^*$ for $i=u,c,t$. The explicit expressions for $\Gamma_{12}^{cc,uu,uc}$ could be found in Refs.~\cite{Lenz,beneke}. It should be noted that $\Gamma_{12}^s$ is dominated by the CKM-favored tree-level $b\to c\bar{c}s$ transition within the SM, and the NP effects are generally negligible~\cite{Lenz}. Hence $\Gamma_{12}^s=\Gamma_{12}^s(SM)$ holds as a good approximation, which will be assumed throughout this paper.

\section{Numerical results and discussions}
\label{sec:results}

We now proceed to present our numerical results and discussions. Values of the relevant input parameters used throughout this paper are summarized in Table~\ref{inputs}. Our SM predictions for the observables of $B_s^0-\bar{B}_s^0$ mixing are given in the third row of Table~\ref{pred}, in which the experimental data averaged by the HFAG~\cite{HFAG} are also listed in the second row for comparison. As mentioned already in the introduction section, there is no significant deviation between the SM predictions and the experimental data for the observables at the current level of precision, even through a bit disagreement appears for $a_{sl}^s$. Therefore, these observables are expected to put strong constraints on the parameter spaces of 2HDMs with MFV.

%%%%%%%%%%%%%%%%%%%%%%%%%%%%%%%%%%%%%%%%
\begin{table}[t]
\begin{center}
 \caption{\label{inputs} \small Values of the relevant input parameters throughout this paper.}
 \small
 \vspace{0.2cm}
 \doublerulesep 0.8pt \tabcolsep 0.05in
 \begin{tabular}{l}
 \hline\hline
 $|V_{us}|=0.2253\pm0.0008$\,,\,
 $|V_{ub}|=0.00413\pm0.00049$\,,\,
 $|V_{cb}|=0.0411\pm0.0013$\,,\,
 $\gamma=(68.0^{+8.0}_{-8.5})^{\circ}$\, \cite{PDG14}
 \\ \hline
 $\bar{m}_{s}(2\,{\rm GeV})=95 \pm 5$~MeV\,,\quad
 $\bar{m}_{b}(\bar{m}_{b})=4.18 \pm 0.03$~GeV\,,\quad
 $m_t=173.21\pm 0.87$ GeV\, \cite{PDG14}\\
 $\frac{\bar{m}_s(\mu)}{\bar{m}_{u,d}(\mu)}=27.5 \pm 1.0$ \cite{PDG14}\,,\quad
 $m_{b}^{\rm pow}=4.8^{+0.0}_{-0.2}$ GeV\, \cite{Lenz}
 \\ \hline
 $f_{B_s}=228 \pm5\pm6$~MeV\,,\quad
 $f_{B_s}\sqrt{B_1}=211\pm5\pm6$~MeV\,,\quad
 $f_{B_s}\sqrt{B_2}=195\pm5\pm5$~MeV\,,\\
 $f_{B_s}\sqrt{B_3}=215\pm14\pm9$~MeV\, \cite{Carrasco:2013zta} \\
 \hline \hline
\end{tabular}
\end{center}
\end{table}
%%%%%%%%%%%%%%%%%%%%%%%%%%%%%%%%%%%%%%%%

%%%%%%%%%%%%%%%%%%%%%%%%%%%%%%%%%%%%%%%%
\begin{table}[t]
\begin{center}
 \caption{\label{pred} \small Numerical results for $\Delta M_s[{\rm ps^{-1}}]$, $\Delta \Gamma_s[{\rm ps^{-1}}]$, $\phi^{c\bar{c}s}_s$ and $a_{sl}^s[\%]$ within the SM. The theoretical uncertainties are obtained by varying each input parameter listed in Table~\ref{inputs} within its respective allowed range and then adding the individual uncertainty in quadrature.}
 \small
 \vspace{0.2cm}
 \doublerulesep 0.8pt \tabcolsep 0.25in
 \begin{tabular}{lccccccccccc} \hline \hline
 &    $\Delta M_s$       &     $\phi^{c\bar{c}s}_s$    &$\Delta \Gamma_s$        & $a_{sl}^s[\%]$  \\ \hline
 Exp. & $17.757\pm0.021$ & $-0.015\pm0.035$ & $0.081\pm 0.006$ & $-0.75\pm0.41$ \\
 SM   & $17.228^{+1.731}_{-1.672}$ & $-0.043^{+0.006}_{-0.006}$ & $0.082^{+0.009}_{-0.013}$
      & $0.0026^{+0.0004}_{-0.0004}$\\
 \hline \hline
\end{tabular}
\end{center}
\end{table}
%%%%%%%%%%%%%%%%%%%%%%%%%%%%%%%%%%%%%%%%

From the analytic expressions of the charged-Higgs contributions to the $B_s^0-\bar{B}_s^0$ mixing calculated in section~2.2, one can find that the model parameters relevant to our study include the two Yukawa coupling parameters $A_u$ and $A_d$, as well as the charged-Higgs mass $m_{H^{\pm}}$. For the case of complex couplings, one could equivalently choose $|A_u|$ and $A_dA_u^{*}=|A_dA_u^{*}|e^{-i\theta}$ as the independent parameters, with $\theta$ being the relative phase between the two Yukawa coupling parameters. For the parameter $|A_u|$, as detailed in Ref.~\cite{Slavich}, an upper bound can be obtained from the $Z\to b\bar{b}$ decay. The parameter $A_d$ is, however, much less constrained phenomenologically~\cite{Slavich,Yuan}. Concerning the charged-Higgs mass, we shall use the LEP lower bound $m_H^{\pm}>78.6$~GeV ($95\%$ CL)~\cite{LEPHpm}, which is obtained under the assumption that $H^{\pm}$ decays dominantly into fermions and does not refer to any specific Yukawa structure. Direct searches for charged-Higgs bosons are also performed by the Tevatron~\cite{Gutierrez:2010zz}, ATLAS~\cite{ATLASHpm} and CMS~\cite{CMSHpm} collaborations. However, most of the limits on $m_H^{\pm}$ depend strongly on the assumed Yukawa structure. In this paper, we shall generate randomly numerical points for the model parameters in the ranges
\begin{align}\label{eq:range}
|A_u|\in[0,3]\,,\qquad m_H^{\pm}\in[80,500]~\mathrm{GeV}\,,
\end{align}
whereas no severe constraints for $|A_dA_u^{*}|$~(or $|A_d|$) and $\theta$.

To compare the relative strength of the charged-Higgs contributions with respect to the SM one at the scale $\mu_b=m_b$, choosing $m_{H^{\pm}}=200$~GeV and the default values of the input parameters listed in Table~\ref{inputs}, we get
\begin{align}
\label{eq:AIII1}
\frac{{\cal A}^{VLL}}{(V_{tb}V_{ts}^*)^2}\times 10^{8}\simeq&3.73+2.00|A_u|^2 + 0.45|A_u|^4\,,\\[0.2cm]
\label{eq:AIII2}
\frac{{\cal A}^{SRR}}{(V_{tb}V_{ts}^*)^2}\times 10^{12}\simeq &-3.21|A_u|^2+0.69|A_u|^4+5.99A_dA_u^{*}-3.44|A_u|^2 A_dA_u^{*}+3.44A_d^2A_u^{*2}\,,\\[0.2cm]
\label{eq:AIII3}
\frac{{\cal A}^{TRR}}{(V_{tb}V_{ts}^*)^2}\times 10^{12}\simeq&0.03|A_u|^2 -0.01|A_u|^4 -0.05A_dA_u^{*} +0.03|A_u|^2 A_dA_u^{*}-0.03A_d^2A_u^{*2}\,,
  \end{align}
in the case of type-III model, and
\begin{align}
\label{eq:AC1}
\frac{{\cal A}^{VLL}}{(V_{tb}V_{ts}^*)^2}\times 10^{8}\simeq&3.73+0.67|A_u|^2 + 0.28|A_u|^4\,,\\[0.2cm]
\label{eq:AC2}
\frac{{\cal A}^{SRR}}{(V_{tb}V_{ts}^*)^2}\times 10^{12}\simeq &1.10|A_u|^2-0.12|A_u|^4-2.05A_dA_u^{*}+ 0.61|A_u|^2 A_dA_u^{*}-0.61A_d^2A_u^{*2}\,,\\[0.2cm]
\label{eq:AC3}
\frac{{\cal A}^{TRR}}{(V_{tb}V_{ts}^*)^2}\times 10^{12}\simeq&-0.15|A_u|^2 +0.04 |A_u|^4 + 0.28A_dA_u^{*} -0.18|A_u|^2 A_dA_u^{*}+0.18A_d^2A_u^{*2}\,,
  \end{align}
in the case of type-C model, respectively. The number $3.73$ in Eqs.~\eqref{eq:AIII1} and \eqref{eq:AC1} is the SM contribution, while the SM contributions to ${\cal A}^{SRR}$ and ${\cal A}^{TRR}$ are suppressed by the factor $x_b$, making them numerically smaller by about three orders than ${\cal A}^{VLL}_{SM}$ and hence negligible. From the above numerical results, we make the following observations:
\begin{enumerate}
\item[ (i) ] In both the type-III and type-C models, the charged-Higgs contributions to ${\cal A}^{VLL}$ (Eqs.~\eqref{eq:AIII1} and \eqref{eq:AC1}) depend only on the Yukawa coupling parameter $A_u$ via $|A_u|$, and hence are always constructive to the SM one. For a value $|A_u|\sim{\cal O}(1)$, the type-III contribution could be comparable with the SM one, while the type-C model provides a relatively smaller correction.

\item[ (ii) ] Comparing Eqs.~\eqref{eq:AIII2}-\eqref{eq:AIII3} with \eqref{eq:AIII1} (for the type-III model) and Eqs.~\eqref{eq:AC2}-\eqref{eq:AC3} with \eqref{eq:AC1} (for the type-C model), one can see that the NP contributions to ${\cal A}^{SRR}$ and ${\cal A}^{TRR}$ are much smaller than to ${\cal A}^{VLL}$, especially when $|A_d|\sim|A_u|$. This is because the Wilson coefficients $C^{SRR}(\mu_W)$ and $C^{TRR}(\mu_W)$ are always suppressed by the factor $x_b$ with respect to $C^{VLL}(\mu_W)$, both within the SM and in the 2HDMs with MFV.

\item[ (iii) ] In the case with large complex values of $A_dA_u^{*}$, however, the charged-Higgs contributions to ${\cal A}^{SRR}$ and ${\cal A}^{TRR}$ could provide a large imaginary part to the off-diagonal mass matrix element $M_{12}^s$, which may result in a significant correction to the CP-violating observables, such as $\phi_s$ and $\phi_s^{c\bar{c}s}$.

\item[ (iv) ] Different from the type-C model, the type-III contribution to ${\cal A}^{TRR}$ is induced only by the RG evolution effect, and is numerically much smaller. There are, however, cancelations between the charged-Higgs contributions to ${\cal A}^{SRR}$ and ${\cal A}^{TRR}$ in the type-C model.
\end{enumerate}
It is therefore expected that the current experimental data on $B_s^0-\bar{B}_s^0$ mixing could put some constraints on the model parameters and be used to distinguish between these two models.

\begin{figure}[t]
\begin{center}
  \subfigure[]{\includegraphics[width=7.8cm]{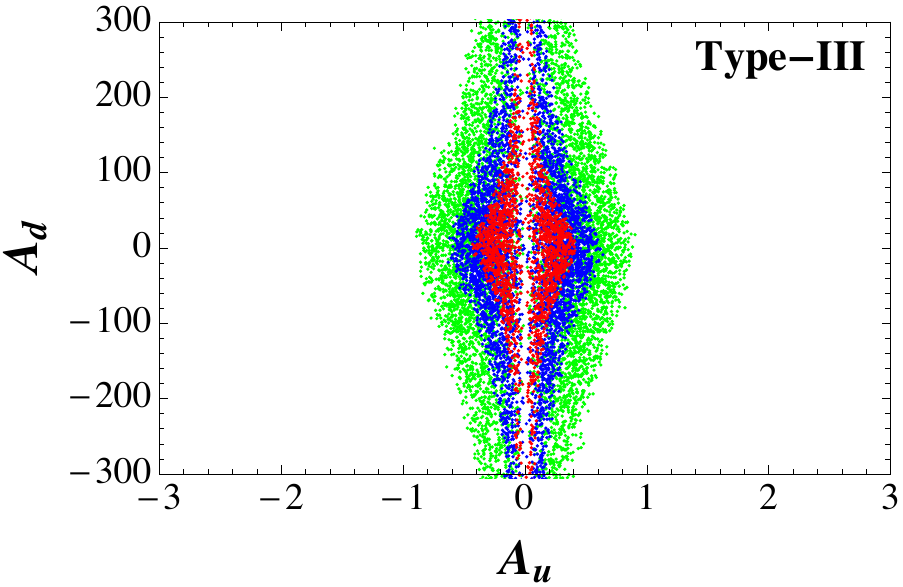}} \qquad
  \subfigure[]{\includegraphics[width=7.8cm]{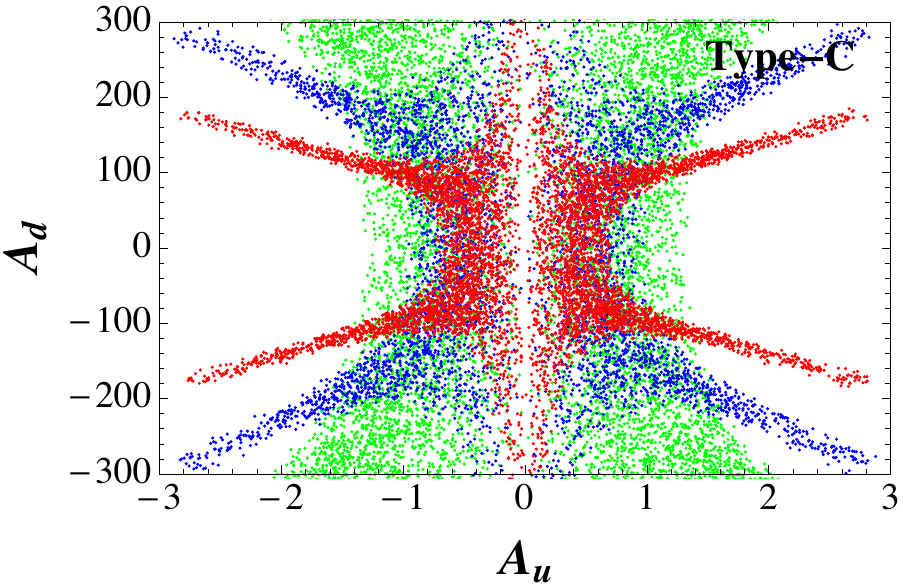}}
  \caption{\small Allowed spaces of the parameters $A_u$ and $A_d$ in type-III and type-C models under the combined constraints from $\Delta M_s$, $\phi^{c\bar{c}s}_s$ and $a_{sl}^s$, in the case of real couplings. The red, blue and green pointed regions are obtained with $m_{H^{\pm}}=100$, $250$ and $500~{\rm GeV}$, respectively.}
  \label{fig:realcase}
\end{center}
\end{figure}

\begin{figure}[t]
\begin{center}
  \subfigure[]{\includegraphics[width=7.8cm]{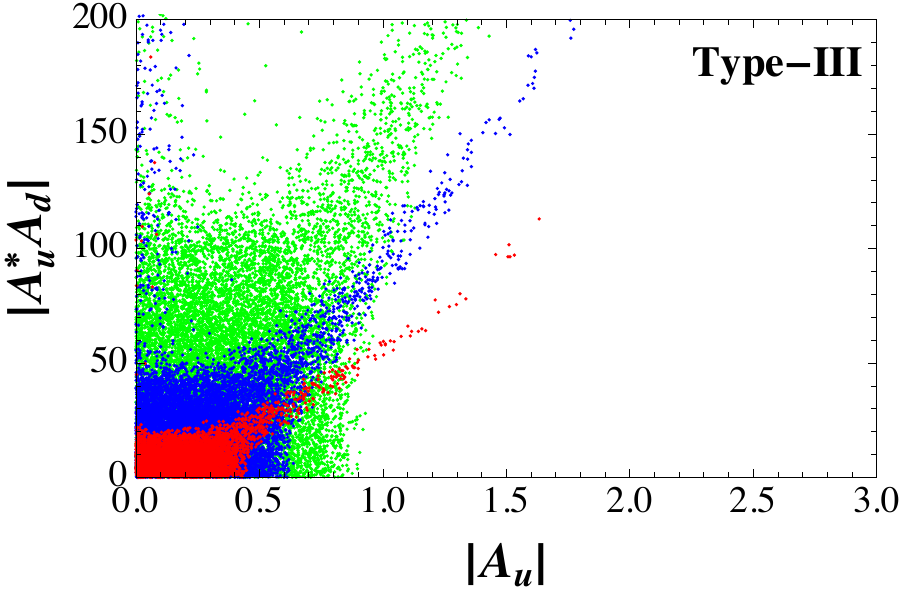}} \qquad
  \subfigure[]{\includegraphics[width=7.8cm]{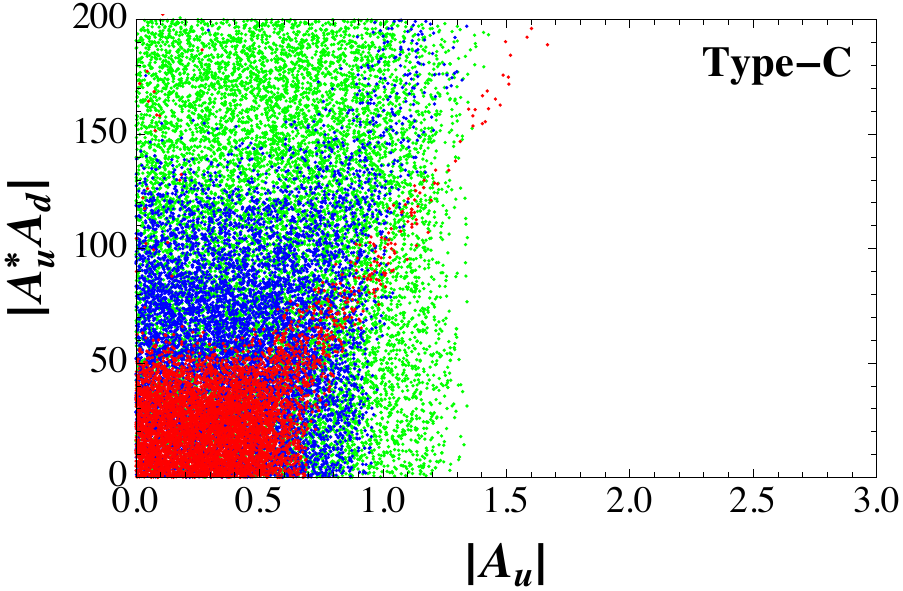}}\\
  \subfigure[]{\includegraphics[width=7.8cm]{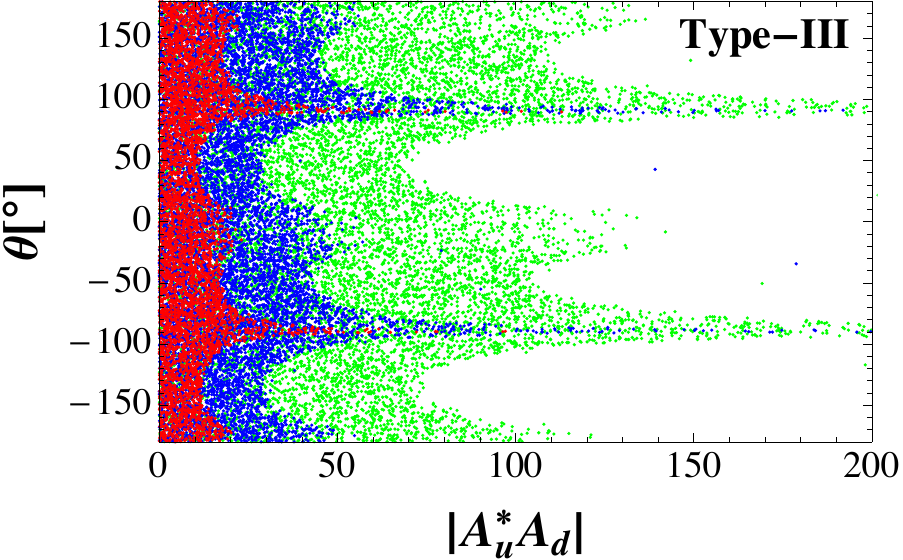}} \qquad
  \subfigure[]{\includegraphics[width=7.8cm]{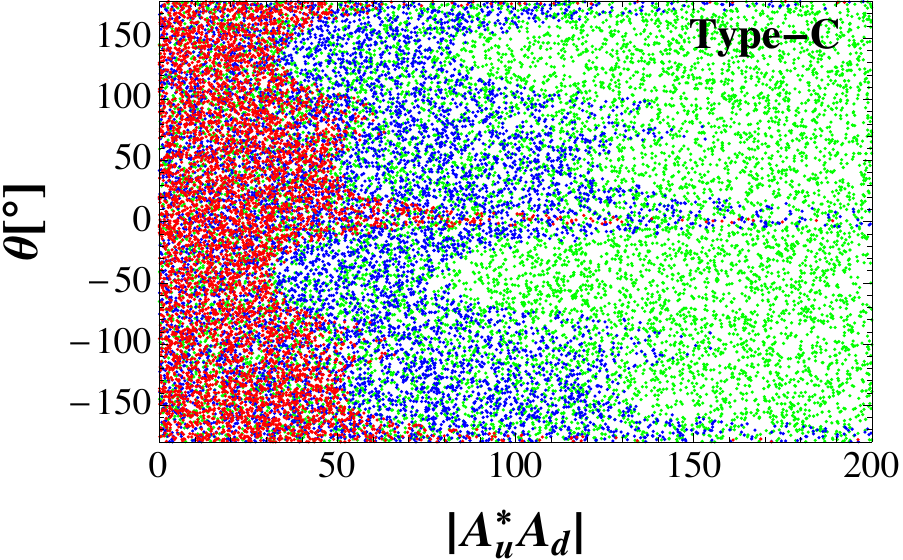}}
  \caption{\small Allowed spaces of the parameters $|A_u|$, $|A_u^*A_d|$ and $\theta$ in type-III and type-C models under the combined constraints from $\Delta M_s$, $\phi^{c\bar{c}s}_s$ and $a_{sl}^s$, in the case of complex couplings. The other captions are the same as in Fig.~\ref{fig:realcase}.}
  \label{fig:comcase}
\end{center}
\end{figure}

To get the explicitly allowed parameter spaces, we perform the analysis with the following procedure: we scan the parameter spaces within the ranges specified by Eq.~\eqref{eq:range}, with the value of $m_{H^{\pm}}$ fixed at $100$, $250$ and $500~{\rm GeV}$, respectively. At each point in the parameter spaces, we evaluate the theoretical prediction for an observable, together with the corresponding theoretical uncertainty induced by the input parameters listed in Table~\ref{inputs}. The theoretical range for an observable at each point is obtained by varying each input parameter within its respective allowed range and then adding the individual uncertainty in quadrature. If the obtained theoretical range has overlap with the $2\sigma$ range of the experimental data, the point is regarded as allowed. In addition, we consider two different cases: real and complex couplings with respect to $A_u$ and $A_d$. Under the combined constraints from $\Delta M_s$, $\phi^{c\bar{c}s}_s$ and $a_{sl}^s$, the allowed parameter spaces of the type-III and type-C models are shown in Figs.~\ref{fig:realcase} (for the case of real couplings) and \ref{fig:comcase} (for the case of complex couplings), respectively.

%%%%%%%%%%%%%%%%%%%%%%%%%%%%%%%%%%%%%%%%
%%%%%%%%%%%%%%%%%%%%%%%%%%%%%%%%%%%%%%%%

%%%%%%%%%%%%%%%%%%%%%%%%%%%%%%%%%%%%%%%%
%%%%%%%%%%%%%%%%%%%%%%%%%%%%%%%%%%%%%%%%

For the case of real couplings, it can be seen from Fig.~\ref{fig:realcase} that:
\begin{enumerate}
\item[ (i) ] In the type-III model, as shown in Fig.~\ref{fig:realcase}(a), the module of Yukawa coupling parameter $A_u$ is severely constrained by the good agreement between the SM prediction and the experimental data for $\Delta M_s$; for instance $|A_u|<1$ is obtained with $m_{H^{\pm}}=500\,{\rm GeV}$. There are, however, almost no constraints on the coupling $A_d$, because the contribution involving it is negligible with respect to the one involving only $A_u$.

\item[ (ii) ] In the type-C model, because the charged-Higgs contribution to ${\cal A}^{VLL}$ is relatively small and large cancelation effects exist between the terms involving $A_d$ and $A_u$, the allowed values of $A_d$ and $A_u$ could be large simultaneously, with either the same or the opposite signs, as shown by the four ``legs" in Fig.~\ref{fig:realcase}(b).

\item[ (iii) ] Besides the ``legs" in Fig.~\ref{fig:realcase}(b), the difference between the two models is also featured by the different shapes of the allowed parameter spaces. The current data on $B_s^0-\bar{B}_s^0$ mixing generally puts a stronger constraint on the type-III model; for instance, with the assumption $|A_d|\sim |A_u|$ and choosing $m_{H^{\pm}}=500\,{\rm GeV}$, the upper bound $|A_u|\sim 1.5$ obtained in type-C model is obviously looser than the one $|A_u|\sim 1$ in type-III model.
\end{enumerate}

For the case of complex couplings, one more model parameter $\theta$ is introduced. From Fig.~\ref{fig:comcase}, it is found that:
\begin{enumerate}
\item[ (i) ] In the type-III model, as shown in Figs.~\ref{fig:comcase}(a) and \ref{fig:comcase}(c), large values of $|A_u|$ and $|A_u^*A_d|$ are still allowed around $\theta\sim \pm90^{\circ}$, which makes it different from the case of real couplings. This is due to the fact that large cancelation effects appear among the charged-Higgs contributions when $\theta\sim \pm90^{\circ}$, which can also be seen from Eqs.~\eqref{eq:AIII2} and \eqref{eq:AIII3}. Moreover, as shown in Fig.~\ref{fig:comcase}(a), an approximately linear relationship is observed between $|A_u^*A_d|$ and $|A_u|$ when $|A_u|\gtrsim0.5$.

\item[ (ii) ] As shown in Figs.~\ref{fig:comcase}(b) and \ref{fig:comcase}(d), similar observations could also be made in the type-C model, except for the fact that the constraints on the model parameters are now much looser. In addition, the cancelation effects among the charged-Higgs contributions occur around $\theta\sim 0^{\circ}$ and $\pm180^{\circ}$, which is different from that observed in the type-III model.

\end{enumerate}

From the above discussions, we conclude that the type-III and type-C models exhibit some significantly different behaviors under the experimental constraints from $B_s^0-\bar{B}_s^0$ mixing. However, due to the large theoretical and experimental uncertainties, the differences in the small $|A_u|$ and $|A_d|$ ranges are hardly to be distinguished from each other. The future refined measurement and precise theoretical evaluation for $B_s^0-\bar{B}_s^0$ mixing might show a much clearer phenomenological picture for the type-III and type-C models.

As a final comment, it should be noted that the same analysis could also be applied to the $B_{d}^0-\bar{B}_d^0$ mixing, which is another important related low-energy process. The charged-Higgs effect on it can be obtained from that on the $B_{s}^0-\bar{B}_s^0$ mixing, with the replacement $s\to d$ throughout the theoretical formulae presented in Sec.~\ref{subsec:wcs}. However, we find that the bounds on the model parameters derive from the $B_{d}^0-\bar{B}_d^0$ mixing are quite similar to the ones from the $B_{s}^0-\bar{B}_s^0$ mixing, and no any further information on the model parameters could be obtained from the former. Therefore, the constraints from $B_{d}^0-\bar{B}_d^0$ mixing will not be shown here.

\section{Conclusion}
\label{sec:conc}

In this paper, we have calculated the one-loop electro-weak corrections to the $B_s^0-\bar{B}_s^0$ mixing within the type-III and type-C 2HDMs with MFV, in which the second scalar doublet is fixed to be color-singlet and color-octet, respectively. It is noted that, in order to get a gauge-independent result, the external momenta of the heavy quarks inside the mesons should be taken into account, and the heavy-quark masses should be kept up to the second order.

Based on the obtained short-distance Wilson coefficients of the four-quark operators appearing in the effective weak Hamiltonian, and combining the up-to-date experimental data on $B_s^0-\bar{B}_s^0$ mixing, we then performed a detailed phenomenological analysis of the charged-Higgs effects on this process. Our main conclusions are summarized as follows:
\begin{enumerate}
\item[ (i) ] While the type-C model gives a nonzero contribution to the Wilson coefficient $C^{TRR}_{C}(\mu_W)$, the type-III contribution to the amplitude ${\cal A}^{TRR}$ is induced only by RG evolution effect.

\item[ (ii) ] In the case of real couplings, the allowed spaces of the Yukawa coupling parameters $A_u$ and $A_d$ in the two models are obviously different, as shown in Fig.~\ref{fig:realcase}.

\item[ (iii) ] In the case of complex couplings, due to the cancelation effects among the charged-Higgs contributions, large values of $|A_u|$ and $|A_d|$ are still allowed around $\theta\sim \pm 90^{\circ}$ in the type-III and around $\theta\sim0^{\circ}\,,\pm180^{\circ}$ in the type-C model, which is shown in Fig.~\ref{fig:comcase}.
\end{enumerate}
The observed differences could be used to distinguish the two models. It should be noted, however, that their differences in the small $|A_d|$ and $|A_u|$ ranges are hardly to be distinguished, due to the large theoretical and experimental uncertainties. More refined theoretical and experimental efforts are therefore needed for a much clearer phenomenological picture.

\section*{Acknowledgments}

We thank Antonio Pich and Martin Jung for useful discussions and cross-checks for the box Feynman diagrams. This work is supported by the National Natural Science Foundation of China (Grant Nos. 11475055, 11435003, 11105043 and 11005032). Q.~Chang is also supported by the Foundation for the Author of National Excellent Doctoral Dissertation of P.~R.~China~(Grant No.~201317) and the Program for Science and Technology Innovation Talents in Universities of Henan Province~(No.~14HASTIT036). X.~Q.~Li is also supported by the Scientific Research Foundation for the Returned Overseas Chinese Scholars, State Education Ministry, by the Open Project Program of SKLTP, ITP, CAS, China~(No.~Y4KF081CJ1), and by the self-determined research funds of CCNU from the colleges' basic research and operation of MOE~(CCNU15A02037).

\begin{appendix}

\section{Relevant coefficients for $B_s^0-\bar{B}_s^0$ mixing}
\label{appendix:1}

Here we present the explicit expressions for $f_i(x_t,x_h)$ appearing in Eqs.~\eqref{eq:Ci_III} and \eqref{eq:Ci_C}:
\begin{align} \label{fi_def}
f_1(x_t,x_h) & = \frac{1}{2}\left[\frac{x_t^2(x_t-4)}{(x_h-x_t)(1-x_t)} +\frac{x_t^2(3x_t^2-x_h(4-2x_t+x_t^2))\ln{x_t}}{(x_h-x_t)^2(1-x_t)^2}
-\frac{(x_h-4)x_hx_t^2\ln{x_h}}{(1-x_h)(x_h-x_t)^2}\right]\,,\\[0.2cm]
f_2(x_t,x_h) & = \frac{1}{2}\left[\frac{x_t^2(x_h+x_t)}{2(x_h-x_t)^2} +\frac{x_hx_t^3\ln{x_t}}{(x_h-x_t)^3}-\frac{x_hx_t^3\ln{x_h}}{(x_h-x_t)^3}\right]\,,\\[0.2cm]
f_3(x_t,x_h) & = \frac{1}{3} \bigg[-\frac{x_t^2(x_t^2+x_h^4)(-11+7x_t-2x_t^2)+x_hx_t^3(7+53x_t-55x_t^2+19x_t^3)}{3(1-x_h)^2(x_h-x_t)^3(1-x_t)^3}\nonumber\\[0.1cm]
&\hspace{1.1cm}-\frac{x_h^2x_t^2(-2-55x_t+15x_t^2+17x_t^3-11x_t^4)+x_h^3x_t^2(19+17x_t-19x_t^2+7x_t^3)}{3(1-x_h)^2(x_h-x_t)^3(1-x_t)^3}\nonumber\\[0.1cm]
&\hspace{1.1cm}+\frac{2x_t^2(x_h^3-3x_h^2x_t+3x_hx_t^2-3x_t^4+3x_t^5-x_t^6)\ln{x_t}}{(x_h-x_t)^4(1-x_t)^4}\nonumber\\[0.1cm]
&\hspace{1.1cm}-\frac{2x_hx_t^2(x_h^2+(-3+x_h)x_hx_t+(3+(-3+x_h)x_h)x_t^2)\ln{x_h}}{(1-x_h)^3(x_h-x_t)^4}\bigg]\,,\\[0.2cm]
f_4(x_t,x_h) & = -\frac{x_t^2((x_h^2+x_t)(-3+x_t)+x_h(1+6x_t-3x_t^2))}{2(1-x_h)(x_h-x_t)^2(1-x_t)^2}\nonumber\\[0.1cm]
&\hspace{0.46cm}-\frac{x_t^2(x_h^2-2x_hx_t-(-2+x_t)x_t^3)\ln{x_t}}{(x_h-x_t)^3(1-x_t)^3}+\frac{x_hx_t^2(x_h-2x_t+x_hx_t)\ln{x_h}}{(1-x_h)^2(x_h-x_t)^3}\,,\\[0.2cm]
f_5(x_t,x_h) & = -\frac{2x_t^2}{(x_h-x_t)^2}-\frac{x_t^2(x_h+x_t)\ln{x_t}} {(x_h-x_t)^3}+\frac{x_t^2(x_h+x_t)\ln{x_h}}{(x_h-x_t)^3}\,,\\[0.2cm]
f_6(x_t,x_h) & = \frac{1}{6} \bigg[-\frac{x_t^2(5x_h^2-22x_hx_t+5x_t^2)}{3(x_h-x_t)^4} -\frac{x_t^2(x_h^3-3x_h^2x_t-3x_hx_t^2+x_t^3)\ln{x_t}}{(x_h-x_t)^5}\nonumber\\[0.1cm]
&\hspace{1.1cm}+\frac{x_t^2(x_h^3-3x_h^2x_t-3x_hx_t^2+x_t^3)\ln{x_h}}{(x_h-x_t)^5}\bigg]\,,\\[0.2cm]
f_7(x_t,x_h) & = \frac{2x_t^2}{(x_h-x_t)^2}+\frac{x_t^2(x_h+x_t)\ln{x_t}}{(x_h-x_t)^3} -\frac{x_t^2(x_h+x_t)\ln{x_h}}{(x_h-x_t)^3}\,.
\end{align}

\end{appendix}

\end{document}